\documentclass[preprint,12pt]{elsarticle}

\usepackage{lineno}
\usepackage{float}
\usepackage{times}
\usepackage{amsmath,amsthm}
\usepackage{color}
\usepackage{amssymb}
\usepackage{epsfig}

\usepackage{array}
\usepackage{graphicx}
\graphicspath{{figure/}}
\usepackage{subfigure}
\usepackage{epsfig}
\usepackage{color}
\usepackage{flushend}
\usepackage[ruled,vlined,linesnumbered]{algorithm2e}

\newtheorem{theorem}{Theorem}

\newtheorem{corollary}{Corollary}

\newtheorem{lemma}{Lemma}

\begin{document}

\begin{frontmatter}

\title{Game-theoretical approach for task allocation problems with constraints}
\author[label1]{Chunxia Liu}
\author[label2]{Kaihong Lu}
\author[label1]{Xiaojie Chen\corref{cor}}
\cortext[cor]{Corresponding author} \ead{xiaojiechen@uestc.edu.cn}

\author[label3]{Attila Szolnoki}

\address[label1]{School of Mathematical Sciences, University of Electronic Science and Technology of China, Chengdu 611731, China}
\address[label2]{College of Electrical Engineering and Automation, Shandong University of Science and Technology, Qingdao 266590, China}
\address[label3]{Institute of Technical Physics and Materials Science, Centre for Energy Research, H-1525 Budapest P.O. Box 49, Hungary}

\begin{abstract}
The distributed task allocation problem, as one of the most interesting distributed optimization challenges, has received considerable research attention recently. Previous works mainly focused on the task allocation problem in a population of individuals, where there are no constraints for affording task amounts. The latter condition, however, cannot always be hold. In this paper, we study the task allocation problem with constraints of task allocation in a game-theoretical framework. We assume that each individual can afford different amounts of task and the cost function is convex. To investigate the problem in the framework of population games, we construct a potential game and calculate the fitness function for each individual. We prove that when the Nash equilibrium point in the potential game is in the feasible solutions for the limited task allocation problem, the Nash equilibrium point is the unique globally optimal solution. Otherwise, we also derive analytically the unique globally optimal solution. In addition, in order to confirm our theoretical results, we consider the exponential and quadratic forms of cost function for each agent. Two algorithms with the mentioned representative cost functions are proposed to numerically seek the optimal solution to the limited task problems. We further perform Monte Carlo simulations which provide agreeing results with our analytical calculations.
\end{abstract}


\end{frontmatter}


\section{Introduction}
Distributed optimization problems can frequently be detected in engineering~\cite{Balaji_pg10, Pantoja_a12, Pantoja_a19, Mojica-Nava_e17, Pantoja_a11, Obando_g13, Wang_d18,Turner-IEEETC18} including traffic~\cite{Balaji_pg10}, lighting systems~\cite{Pantoja_a12}, multirobot systems~\cite{Turner-IEEETC18}, and generator power systems~\cite{Pantoja_a19, Mojica-Nava_e17, Pantoja_a11, Obando_g13, Wang_d18}. Indeed, the related problems can be solved by minimizing the global objective function in a multi-agent system, where each individual can only obtain the information of its adjacent neighbors in a connected graph~\cite{Liu_q17,Nedic_a17,Shi-IEEETC19,Tian-IEEETC21,Liu-SCL21}.

As one of the most popular distributed optimization problems, the distributed task allocation problem has collected significant research interest in the last decade~\cite{Wang_xf12,Khattab_mm96,Federgruen_a83,Yin_py06,Nedic_a20}. It has been solved by means of different approaches including Gradients~\cite{Cassandras_cg05,Chen_f21}, Distributed Control~\cite{Olfati-Saber_r07,Yi_p16,Xu_j18}, Consensus Algorithm~\cite{Cao_m05,Fang_l05,Tsianos_ki12,Menache_i11,Chen_g21}, and Game Theory~\cite{Pantoja_a12, Pantoja_a19,Jie_y20,wang_q18,Huang_y23,Barreiro-Gomez_j16,Sun-IEEETC19,Jaleel_h20,Sun_c20,Tan_s17,Zhu-IEEETC22,zhu_2022_IEEETNSE,zhu_2023_Automatica}. In particular, Pantoja \emph{et al}. studied the allocation problem by analysing the distributed replicator equation~\cite{Pantoja_a12, Pantoja_a19}, in which each individual has a fitness function and the task is dynamically adjusted based on these functions. With this approach the optimal solution to the problem can be derived. Recently, game-theoretical approach for solving the distributed optimization problems has been received considerable attention~\cite{Jaleel_h20,Ocampo-Martinez-IEEECSM17,Riehl-ARC18,Marden-ARCRAS18,Rizk-IEEETCDS18}.

Notably, previous works mainly focused on the task allocation problem in a number of agents where there are no constraints on the amount of task that each individual affords in a game-theoretical framework~\cite{Barreiro-Gomez_j16,Tan_s17}. It is worth mentioning that there exists the globally optimal solution in this scenario when the cost function is strictly convex~\cite{Pantoja_a12, Pantoja_a19}. Indeed, each individual can only afford a limited amount of task due to their finite capacities or physical strengths or economic factors~\cite{Pantoja_a19,Chen_EPL10}. Thus, it is significant to address the task allocation problem with constraints of task allocation. Note that some works have already studied this research path~\cite{Chen_f21, Yi_p16, Xu_j18,Chen_g21}. For example, Chen et al. considered a distributed gradient-descent algorithm for the distributed optimization problem with equality constraints and local box constraints~\cite{Chen_f21}. In Ref.~\cite{Yi_p16}, an initialization-free distributed algorithm is proposed for the problem with coupled equality constraints and local feasibility constraints. Xu et al. proposed a novel distributed algorithm to solve the above problem based on duality analysis~\cite{Xu_j18}. In Ref.~\cite{Chen_g21}, a distributed surplus-based algorithm is proposed to solve the problem with inequality constraints, and the convergence is analyzed under strongly connected directed graphs. Notably, these works mainly solved the problem via distributed algorithms. However, thus far few works have used game-theoretical approach to study the task allocation problem with constraints of task allocation. Recently, Martinez-Piazuelo considered the problem of GNE seeking in population games under fairly general constraints~\cite{Piazuelo_m22}. However, it is unclear how the optimal solution to the task problem with constraints of task allocation can be obtained in the proposed game-theoretical framework.

Motivated by this scientific challenge, in this paper we study the task allocation problem with constraints of task affording and convex cost function in a game-theoretical framework. We construct a potential game and calculate the fitness function for each individual. We hence derive the unique globally optimal solution. Furthermore, we respectively consider the exponential and quadratic forms of cost function for each agent and present two algorithms to numerically seek the optimal solution to the limited task allocation problem.

The remainder of this article is arranged as follows. Section~II presents our model and method. Section~III presents our theoretical analysis of exploring the globally optimal solution and the uniqueness of the solution for the limited task allocation problem. In Section~IV, we propose two algorithms to numerically seek the globally optimal solution, while numerical examples are provided in Section~V. Finally, conclusions are drawn in Section~VI.

\section{Model and method}
\subsection{Graphs}

We use a graph-theoretical tool to describe the structure of a multi-agent system~\cite{Barreiro-Gomez_j16}. To be specific, let $G=(\nu,\varepsilon)$ be an undirected  connected topology, where $\nu=\{1,2,\cdots,n\}$ denotes the set of nodes representing individuals in the population and $\{(i,j)|i,j\in \nu\} \in \varepsilon$ indicates that individual $i$ and individual $j$ are mutually connected.
Then, let $A=[a_{ij}]$ be an $n\times n$ adjacency matrix whose elements satisfy the following property
\begin{eqnarray*}
  a_{ij}&=&
  \left\{
  \begin{aligned}
  &1,  \quad  \text{if}\ (i,j)\in \varepsilon,\\
  &0,  \quad  \text{otherwise,}\\
 \end{aligned}
  \right.
\end{eqnarray*}
where $a_{ij}=a_{ji}$ for all $i,j \in \nu$.

We define $N_i=\{j|(i,j)\in \varepsilon\}=\{j|a_{ij}=1, j \in \nu\}$ to be the set of neighbors of a node $i \in \nu$.

\subsection{Distributed optimization problem}

In this paper, we use a graph to depict the population structure in which there are $n$ individuals and assume that the total amount of task for the population of individuals is a fixed value, given by $w$. We assume that the total task is undertaken by these $n$ individuals and suppose that $w_i$ is the task amount which individual $i$ bears. However, each individual has the thresholds of task allocation and the upper and lower limits that individual $i$ can undertake are respectively set as $\overline{w}_{i}$ and $\underline{w}_{i}$. Here, $\underline{w}_i\leq w_i \leq \overline{w}_i$. In addition, we have $\sum_{i=1}^nw_i=w$. Furthermore, individual $i$ needs to pay the cost when it undertakes the task and we assume that the cost function of individual $i$ is given as $c_i(w_i)$.  In this work, we aim to study how to optimally allocate the total task among the $n$ individuals, so that the total cost amount can be minimized. Accordingly, this limited task allocation problem can be written as follows
 \begin{align}\label{key1}
   \begin{split}
	        &{\rm min} \quad C(W)=\sum_{i=1}^nc_i(w_i),\\
            &{ s.t.}  \quad \sum_{i=1}^nw_i=w,\\
            &\qquad \underline{w}_i\leq w_i \leq \overline{w}_i,		
   \end{split}
\end{align}
where $C(W)$ is the total cost function and $W=(w_1,w_2,\cdots,w_n)^T$. Here, we suppose that $C(W)$ is a strictly convex function, so that there exists the optimal solution~\cite{Bertsekas_d03}. Accordingly, the feasible solutions for the limited task allocation problem can be given as
\begin{eqnarray*}
\textit{S}=\{W|\sum_{i=1}^nw_i=w \quad {\rm and} \quad \underline{w}_i\leq w_i \leq \overline{w}_i\}.
\end{eqnarray*}

\section{Theoretical analysis}
\subsection{Preliminaries}

Population games are often used to study the strategic interactions in a large population of
players, which can be used to solve distributed optimization problem of limited task allocation~\cite{Pantoja_a12,Pantoja_a19}. Accordingly, we consider that in a population with size $n$, each individual $i$ can bear an amount of task $w_i$ ($w_i\geq 0$) in a strategic interaction. We then use an $n$ dimension vector $W=(w_1,w_2,\cdots,w_n)^T$ to depict the population state which describes how the total task is completely allocated among all the $n$ individuals. Hence the state distribution in the population can be described by a simplex~\cite{Quijano_n17}, that is,
\begin{eqnarray*}
\triangle&=&\{W\in R_+^n|\sum_{i=1}^nw_i=w\},
\end{eqnarray*}
where $R_+^n=\{W=(w_1,w_2,\cdots,w_n)^T|w_i\geq 0\}$. It is easy to see that $\triangle  \subseteq  \textit{S}$.

In the framework of population games, each individual $i$ who bears the amount of task $w_i$ can obtain a payoff, which is described by a payoff function $f_i(w_i)$. Accordingly, we have $f_i:[\underline{w}_i,\overline{w}_i] \longmapsto R$ and then can have the payoff vector for the population game~\cite{Barreiro-Gomez_j16}, which is given as $$F(W)=(f_1(w_1),f_2(w_2),\cdots,f_n(w_n))^T.$$

In a multi-agent system, we can apply a potential game to describe the optimization problem of the population by choosing an appropriate payoff function~\cite{Pantoja_a19,Sandholm_wh10}. We thus define that the payoff vector $F$ satisfies the following equation
\begin{eqnarray*}
f_i(w_i)=\frac{\partial [-C(W)]}{\partial w_i}=-\frac{\partial C(W)}{\partial w_i}=-\frac{d c_i(w_i)}{d w_i}\,,
\end{eqnarray*}
hence $F$ is a potential game.

According to Refs.~\cite{Barreiro-Gomez_j16,Tan_s17}, we can conclude that there exists a Nash equilibrium $W^*=(w_1^*,w_2^*,\cdots,w_n^*)^T$ in the potential game for the optimization problem~\cite{Sandholm_wh10}. Correspondingly, we can have
\begin{eqnarray*}
\begin{split}
NE(F)=\{W^* \in \triangle|w_i^*>0\Rightarrow f_i(w_i^*) \geq f_j(w_j^*),\forall i,j\in v\}.
\end{split}
\end{eqnarray*}
Therefore, if $W^* \in NE(F) \cap R_{++}^n$, it is easy to see that $f_i(w_i^*)=f_j(w_j^*)$ for all $i,j\in v$, where  $R_{++}^n=\{W=(w_1,w_2,\cdots,w_n)^T|w_i > 0\}$.

\subsection{Globally optimal solution}
Here we study the globally optimal solution in two different cases as follows.

\subsubsection{The case of $W^*\in NE(F)\cap \textit{S}$}

Based on the above description, we state the following conclusion described by Theorem~\ref{th1}.
\begin{theorem}\label{th1}
 If $W^*\in NE(F)\cap \textit{S}$ , then $W^*$ is the globally optimal solution for the limited task allocation problem described by Eq.~\eqref{key1}.
\end{theorem}

\emph{Proof:} The limited task allocation problem described by Eq.~\eqref{key1} can be expressed as
\begin{align}\label{key2}
   \begin{split}
	        &{\rm min} \quad C(W)=\sum_{i=1}^nc_i(w_i),\\
            &{ s.t.} \quad w_i-\underline{w}_i \geq 0,\\
            &\qquad \overline{w}_i-w_i \geq 0,\\
            &\qquad \sum_{i=1}^nw_i-w=0\,.\\
   \end{split}
\end{align}
Here, for the sake of convenience we define $p_i(W)=w_i-\underline{w}_i$, $q_i(W)=\overline{w}_i-w_i$, and $h(W)=
\sum_{i=1}^nw_i-w$. Accordingly, the feasible solutions for the limited task allocation problem can be given as
\begin{eqnarray*}
S=\{W|h(W)=0, p_{i}(W) \geq 0, q_{i}(W) \geq 0, \forall i=1,2,\cdots,n\}.
\end{eqnarray*}
Note that the functions $p_{i}(W)$, $q_{i}(W)$, and $h(W)$ are all linear with ${w_1,w_2,\cdots,w_n}$. Consequently, the Karush-Kuhn-Tucker (KKT) conditions are both necessary and sufficient for the solution of the problem described by Eq.~\eqref{key2}. We now formulate and solve the specific KKT conditions for the limited task allocation problem. To do that, we assume that gradients
are taken as $(\frac{\partial C(W)}{\partial w_1},\frac{\partial C(W)}{\partial w_2},\cdots,\frac{\partial C(W)}{\partial w_n})^T$ and define $I=\{i|p_{i}(W^*)=0\}$ and $J=\{j|q_{j}(W^*)=0\}$.

According to the above description, the Kuhn-Tucker equality for Eq.~\eqref{key2} is
\begin{eqnarray*}
\begin{split}
\nabla C(W^*)-\sum_{i\in I}\alpha_i\nabla p_{i}(W^*)-\sum_{J\in J}\beta_j\nabla q_{j}(W^*)-\gamma\nabla h(W^*)=0.
\end{split}
\end{eqnarray*}

Due to $W^*\in NE(F) \cap \textit{S}  \subseteq NE(F) \cap R_{++}^n$, we have $f_i(w_i^*)=f_j(w_j^*)$ $(i,j=1,2,\cdots,n)$. For simplicity, we set that $f_i(w_i^*)=-\lambda \in \Re$ which is the fitness value for each $i$. We then have
\begin{eqnarray*}
\nabla C(W^*)&=&(-f_1(w_1^*),-f_2(w_2^*),\cdots,-f_n(w_n^*))^T\\
            &=&(\lambda,\lambda,\cdots,\lambda)^T.
\end{eqnarray*}

There should exist a set of parameters $\alpha_i=0$, $\beta_j=0$, and $\gamma=\lambda$, where $i\in I$ and $j\in J$, to make the following equation satisfied~\cite{Nocedal_j06}
\begin{align}\label{key3}
\begin{split}
	\nabla C(W^*)-\sum_{i\in I}\alpha_i\nabla p_{i}(W^*)-\sum_{j\in J}\beta_j\nabla q_{j}(W^*)-\gamma\nabla h(W^*)=0.
\end{split}
\end{align}
It is easy to show that the KKT conditions are satisfied for $W=W^*$. We can thus conclude that $W^*$ is a globally optimal solution for limited task allocation problem~\cite{Nocedal_j06}.

Furthermore, we can obviously have the following corollary:
\begin{corollary}\label{th5}
If $ NE(F)\cap \textit{S}=\phi$, then $W^{*}$ is the unique optimal solution to the limited task allocation problem described by Eq.~\eqref{key1}.
\end{corollary}

In the framework of population games, the amount of task $w_i$ for actor $i$ can be dynamically adjusted based on the above payoff information. According to the distributed replicator dynamics (DRD)~\cite{Quijano_n17} when the cost function is strictly convex~\cite{Pantoja_a12,Pantoja_a19,Barreiro-Gomez_j16} and individuals have not specific constraints, the optimal allocation scheme can be reached finally. Correspondingly, the dynamical equation for the task allocation is given by~\cite{Barreiro-Gomez_j16}
\begin{align}\label{key7}
   \begin{split}
	        \dot{w}_i &= w_i(f_i(w_i)-\bar{f}(W)) \\
            &= w_i(f_i(w_i)\sum_{j\in N_i}\frac{w_j}{w}-\sum_{j\in N_i}{f_j(w_j)\frac{w_j}{w}}) \\
            &=\frac{w_i}{w}(f_i(w_i)\sum_{j\in N_i}w_j-\sum_{j\in N_i}{f_j(w_j)w_j}),	
   \end{split}
\end{align}
where $\bar{f}(W)$ denotes the average payoff of the population, and we have
\begin{align}\label{key8}
\bar{f}(W)=\sum_{j\in N_i}{f_j(w_j)\frac{w_j}{w}}.
\end{align}

We can accordingly have the conclusion, described by Lemma \ref{le1}.

\begin{lemma}\label{le1}
 If $F$ is a Potential Game with strictly convex potential function $C(W)$ and $W(0)\in \textit{S}$, $W(t)$ will asymptotically converge to $W^*$ under DRD.
\end{lemma}
\emph{Proof:} Due to $W^*\in NE(F)$, it is obvious that $W^*=\operatorname*{argmin}_{W\in \textit{S}}C(W)$. We thus construct a Lyapunov function as
\begin{eqnarray*}
  V(W)=C(W)-C(W^*).
\end{eqnarray*}

Then we can easily see that $V(W)\geq 0$, and $V(W)=0$ iff $W=W^*$. Furthermore, we can calculate $\dot V(W)$ as
\begin{eqnarray*}
\begin{split}
 \dot V(W)&=\sum_{i=1}^n\dot c_i(w_i)=\sum_{i=1}^n\frac{dc_i(w_i)}{dw_i}\dot w_i=\nabla C(W)^T\dot W\\
 &=-F^T\dot W,
 \end{split}
\end{eqnarray*}
where $\dot W=(\dot w_1,\dot w_2,\cdots,\dot w_n)^T$.

According to Refs.~\cite{Pantoja_a19, Bauso_d06}, $\dot V(W)$ can be simplified as
\begin{eqnarray*}
\begin{split}
&\dot V(W) = -F^T\dot W\\
          &=-\sum_{i=1}^n f_i(w_i)\dot w_i\\
          &=-\sum_{i=1}^n f_i(w_i)[\frac{w_i}{w}(f_i(w_i)\sum_{j\in N_i}w_j-\sum_{j\in N_i}{f_j(w_j)w_j})]\\
          &=-\frac{1}{w}(\sum_{i=1}^nf_i^2(w_i)w_i\sum_{j\in N_i}w_j-\sum_{i=1}^nf_i(w_i)w_i\sum_{j\in N_i}f_j(w_j)w_j)\\
          &=-\frac{1}{w}\sum_{(i,j)\in\varepsilon}(f_i^2(w_i)w_iw_j+f_j^2(w_j)w_jw_i-f_i(w_i)w_if_j(w_j)w_j-f_j(w_j)w_jf_i(w_i)w_i)\\
          &=-\frac{1}{w}\sum_{(i,j)\in\varepsilon}w_iw_j(f_i(w_i)-f_j(w_j))^2.
\end{split}
\end{eqnarray*}

Therefore, it is easy to see that $\dot V(W)\leq 0$. We further know that if $f_i(w_i)=f_j(w_j)$ for all $(i,j)\in \varepsilon$, $\dot V(W)=0$. In this case, we also have $W=W^*$. Thus $\dot V(W)=0$
iff $W=W^*$. Hence, if $W(0)\in \textit{S}$, $W(t)$ asymptotically converges to $W^*$ under DRD.

\subsubsection{The case of $ NE(F)\cap \textit{S}=\phi$}

Based on the above description, we can state the following theorem.

\begin{theorem}\label{th4}
If $ NE(F)\cap \textit{S}=\phi$ and $\frac{\partial C(W)}{\partial w_i}$ is continuously differentiable and monotonically increasing, then $W^{o}=(w_1^{o}, w_2^{o},\cdots,w_n^{o})^T \in \textit{S}$ is a globally optimal solution to the limited task allocation problem described by Eq.~\eqref{key1}, given as
\begin{eqnarray*}
  w^{o}_i&=&
  \left\{
  \begin{aligned}
  &\underline{w}_i,  \quad  w_i^*<\underline{w}_i,\\
  &w_i^*,  \quad  \underline{w}_i \leq w_i^* \leq \overline{w}_i,\\
  &\overline{w}_i, \quad w_i^*>\overline{w}_i,\\
 \end{aligned}
  \right.
\end{eqnarray*}
where $W^*=(w_1^*,w_2^*,\cdots,w_n^*)^T \in NE(F)$.
\end{theorem}

\emph{Proof: }The limited task allocation problem described by Eq.~\eqref{key1} can be expressed as
\begin{align}\label{key9}
   \begin{split}
	        &{\rm min} \quad C(W)=\sum_{i=1}^nc_i(w_i),\\
            &{ s.t}. \quad w_i-\underline{w}_i \geq 0,\\
            &\qquad \overline{w}_i-w_i \geq 0,\\
            &\qquad \sum_{i=1}^nw_i-w=0.		
   \end{split}
\end{align}
Here, for convenience we define $p_i(W)=w_i-\underline{w}_i$, $q_i(W)=\overline{w}_i-w_i$, and $h(W)=
\sum_{i=1}^nw_i-w$. Accordingly, the feasible solutions to the limited task allocation problem can be given as
\begin{eqnarray*}
S=\{W|h(W)=0, p_{i}(W) \geq 0, q_{i}(W) \geq 0, \forall i=1,2,\cdots,n\}.
\end{eqnarray*}
Note that the functions $p_{i}(W)$, $q_{i}(W)$, and $h(W)$ are all linear in ${w_1,w_2,\cdots,w_n}$. Consequently, the KKT conditions are both necessary and sufficient for the solution of problem described by Eq.~\eqref{key9}. We now formulate and solve
these conditions for the limited task allocation problem. To be specific, we first
divide the set $\{1,2,\cdots,n\}$ into three
$K$, $I$, and $J$ subsets, where
$$K=\{k|\underline{w}_k<w_k^o<\overline{w}_k\}=\{k_1,k_2,\cdots,k_x\},$$
$$I=\{i|p_{i}(W^o)=0\}=\{i_1,i_2,\cdots,i_y\},$$
and
$$J=\{j|q_{j}(W^o)=0\}=\{j_1,j_2,\cdots,j_z\}.$$
Note that $\left| K \right|=x$, $\left| I \right|=y$, and $\left| J \right|=z$, which respectively represent the number of elements in the subset $K$, $I$, and $J$.

Based on the definition of $NE(F)$, we can conclude that $f_i(w_i^*)=f_j(w_j^*)$, when $w^*_i>0$ and $w^*_j>0$. For simplicity, we set that $f_i(w_i^*)=-\lambda\in \Re$ which is the fitness value for $w^*_i>0$. We can then write the gradient of $C(W)$ as
\begin{eqnarray*}
\nabla C(W)=(\frac{\partial C(W)}{\partial w_{k_1}},\frac{\partial C(W)}{\partial w_{k_2}},\cdots,\frac{\partial C(W)}{\partial w_{k_x}},\\ \frac{\partial C(W)}{\partial w_{i_1}},\frac{\partial C(W)}{\partial w_{i_2}},\cdots,\frac{\partial C(W)}{\partial w_{i_y}},\\
\frac{\partial C(W)}{\partial w_{j_1}},\frac{\partial C(W)}{\partial w_{j_2}},\cdots,\frac{\partial C(W)}{\partial w_{j_z}})^T.
\end{eqnarray*}
In particular, we have
\begin{eqnarray*}
 \frac{\partial C(W^{o})}{\partial w_i}&=&
  \left\{
  \begin{aligned}
  &\lambda,  \qquad  i \in K,\\
  &-f_i(\underline{w}_i), \quad i \in I,\\
  &-f_i(\overline{w}_i), \quad i \in J.\\
 \end{aligned}
  \right.
\end{eqnarray*}

According to the above description, the Kuhn-Tucker equality for Eq.~\eqref{key9} is
\begin{eqnarray*}
\begin{split}
 \nabla C(W^o)-\sum_{i\in I}\alpha_i\nabla p_i(W^o)-\sum_{j \in J}\beta_j\nabla q_j(W^o)-\gamma\nabla h(W^o)=0.
\end{split}
\end{eqnarray*}

We can then respectively calculate $\nabla C(W^o)$, $\nabla p_{i}(W^o)$, $\nabla q_{i}(W^o)$, and $\nabla h(W^o)$ as
\begin{eqnarray*}
\nabla C(W^o)=(\lambda,\cdots,\lambda,
-f_{i_1}(\underline{w}_{i_1}),\cdots,-f_{i_y}(\underline{w}_{i_y}),\\
-f_{j_1}(\overline{w}_{j_1}),-f_{j_2}(\overline{w}_{j_2}),\cdots,-f_{j_z}(\overline{w}_{j_z}))^T,
\end{eqnarray*}
\begin{eqnarray*}
\nabla p_{i}(W^o)=(0,\cdots,0,1,0,\cdots,0)^T  \quad \forall i=i_1,i_2,\cdots,i_y,
\end{eqnarray*}
\begin{eqnarray*}
\nabla q_{j}(W^o)=(0,\cdots,0,-1,0,\cdots,0)^T  \quad \forall j=j_1,j_2,\cdots,j_z,
\end{eqnarray*}
and
\begin{eqnarray*}
\nabla h(W^o)=(1,1,\cdots,1)^T,
\end{eqnarray*}
where the $(x+m)$th element value of $\nabla p_{i_m}(W^o)$ is $1$, other element value of $\nabla p_{i_m}(W^o)$ is $0$; the $(x+y+l)$th element value of $\nabla q_{j_l}(W^o)$ is $-1$, other element value of $\nabla q_{j_l}(W^o)$ is $0$. Here, $m$ ($1\leq m\leq y$) is a positive integer and $l$ ($1\leq l \leq z$) is also a positive integer.

Hence, there should exist a set of parameters $\alpha_i=-f_i(\underline{w}_i)-\lambda$, $\beta_j=\lambda+f_i(\overline{w}_i)$, and $\gamma=\lambda$, where $i\in I$ and $j\in J$, which make the following equation satisfied~\cite{Nocedal_j06}
\begin{align}\label{key10}
\begin{split}
	\nabla C(W^o)-\sum_{i\in I}\alpha_i\nabla p_i(W^o)-\sum_{j \in J}\beta_j\nabla q_j(W^o)-\gamma\nabla h(W^o)=0.
\end{split}
\end{align}
Note that $f_i(w_i)$ is continuously differentiable and monotonically decreasing, we have that $\alpha_i \geq 0$ for $i\in I$ and $\beta_j \geq 0$ for $j\in J$.

Based on the above description, so it is easy to show that the KKT conditions are satisfied for $W=W^o$. We can thus conclude that $W^o$ is a globally optimal solution to the limited task allocation problem~\cite{Nocedal_j06}.

Accordingly, we can obviously obtain the following corollary.
\begin{corollary}\label{th5}
If $ NE(F)\cap \textit{S}=\phi$, then $W^{o}$ is the unique optimal solution to the limited task allocation problem described by Eq.~\eqref{key1}.
\end{corollary}

\section{Seeking solutions}

In this section, the cost functions in realistically limited task allocation problems often has specific forms. In particular, the exponential function~\cite{Chen_x19} and quadratic function~\cite{Pantoja_a12, Pantoja_a19}, which are also typically nonlinear functions, are two of the most commonly used functions to describe the emerging cost. Hence, we present how to search numerically the globally optimal solution to the limited task allocation problem with exponential or quadratic cost function.

\subsection{The case of $W^*\in NE(F)\cap \textit{S}$}

According to Eq.~\eqref{key7}, we can write the following discrete-time dynamical equation for depicting how each individual updates its task allocation amount as
 \begin{align}\label{key11}
\begin{split}
w_i(t+1)=w_i(t)+\Delta t(\frac{w_i(t)}{w}(f_i(w_i(t))\sum_{j\in N_i}w_j(t)-\sum_{j\in N_i}{f_j(w_j(t))w_j(t)})),
\end{split}
\end{align}
where $\Delta t \in \Re^+$ is the fixed step size of the discretization~\cite{Sun_c17,Martinez-Piazuelo_J22}. Here, $w_i(t)$ corresponds to the value of $w_i$ at time $t$. Based on this equation, we can obtain the globally optimal solution numerically.

\subsection{The case of $ NE(F)\cap \textit{S}=\phi$}

\subsubsection{The cost function is exponential}
We consider two specific forms of convex cost functions for $c_i(w_i)$. First, we consider the exponential function
by following previous works~\cite{Chen_x19}, given as,
\begin{equation}\label{exp}
  c_i(w_i)=a_ie^\frac{w_i-\underline{w}_i}{\overline{w}_i-\underline{w}_i},
\end{equation}
where $a_i$ is a positive coefficient factor.

According to Theorem~\ref{th4}, we solve the globally optimal solution to the limited task allocation problem by analysing the Nash equilibrium point. Based on the definition of $NE(F)$, we can conclude that $f_i(w_i^*)=f_j(w_j^*)$, when $w^*_i>0$ and $w^*_j>0$. For simplicity, we set that $f_i(w_i^*)=-\lambda\in \Re$ which is the fitness value for $w^*_i>0$.
Since the cost function is defined by Eq.~\eqref{exp}, we have
\begin{eqnarray*}
 \lambda=\frac{a_i}{\overline{w}_i-\underline{w}_i}e^\frac{w_i^*-\underline{w}_i}{\overline{w}_i-\underline{w}_i}
\end{eqnarray*}
and
\begin{eqnarray*}
w_i^*-\underline{w}_i=(\overline{w}_i-\underline{w}_i)(\ln{\frac{\lambda}{a_i}}+\ln{(\overline{w}_i-\underline{w}_i)}).
\end{eqnarray*}

Furthermore, we have
\begin{eqnarray*}
\sum_{i\in\mathcal{B}}w_i^*-\sum_{i\in\mathcal{B}}\underline{w}_i=\sum_{i\in\mathcal{B}}{u_i^*\ln{\lambda}}+\sum_{i\in\mathcal{B}}{u_i^*\ln{u_i^*}}-\sum_{i\in\mathcal{B}}{u_i^*\ln{a_i}},
\end{eqnarray*}
where $u_i^*=\overline{w}_i-\underline{w}_i$ and $\mathcal{B}=\{i|w^*_i>0, i \in \nu\}$.

Since $w=\sum_{i=1}^nw_i^*=\sum_{i\in\mathcal{B}}w_i^*$, we have
\begin{eqnarray*}
w=a\ln{\lambda}+b,
\end{eqnarray*}
where $a=\sum_{i\in\mathcal{B}}{u_i^*}$ and $b=\sum_{i\in\mathcal{B}}{u_i^*\ln u_i^*}-\sum_{i\in\mathcal{B}}{u_i^*\ln{a_i}}+\sum_{i\in\mathcal{B}}\underline{w}_i$ are constant.

Hence, we can get a linear relationship between $\sum_{i=1}^nw_i$ and $\ln{\lambda}$. Inspired by non-iterative $\lambda$-logic
based algorithm in Ref.~\cite{Sydulu_m99}, we design allocated tasks algorithm for exponential cost function, which can be used for seeking the globally optimal solution. For any undirected connected topology, the details for seeking the solution is presented in Algorithm~\ref{algo:ATA1}. In addition, we know that the computation complexity is about $o(n^2(T_1+T_2))$.
\begin{algorithm}\label{algo:ATA1}
 \caption{Algorithm of task allocation for exponential cost function}
 \label{algo:fa}
 \SetCommentSty{small}
 \tiny
 \KwIn{Total tasks $w$, the number of individual $n$, the upper of tasks $\overline{W}=(\overline{w}_1,\cdots,\overline{w}_n)$, the lower of tasks $\underline{W}=(\underline{w}_1,\cdots,\underline{w}_n)^T$, time scale $T_1$ and $T_2$ , initial value $W^*=(w_1(0),\cdots,w_n(0))$, coefficient factor $a=(a_1,\cdots,a_n)^T$, $n\times n$ adjacency matrix $A=[a_{ij}]$}
 \KwOut{Allocated tasks $W=(w_1, w_2,\cdots,w_n)^T$}
 \LinesNumbered
 \For{each $t$ in range $T_1$}{
    \For{each $i$ in range $n$}{
        $w_i(t+1)=w_i(t)+\Delta t(\frac{w_i(t)}{w}f_i(w_i(t))(\sum_{j\in N_i}w_j(t)
        -\sum_{j\in N_i}{f_j(w_j(t))w_j(t)}))$;
    }
 }
 $W^* = (w_1(T),\cdots,w_n(T))^T$\\
 \For{each $i$ in range $n$}{
    $\lambda_{i\rm min}=\frac{dc_i(\underline{w}_i)}{d\underline{w}_i}$;\\
    $Append(L,\ln\lambda_{i\rm min})$;\\
    $\lambda_{i\rm max}=\frac{dc_i(\overline{w}_i)}{d\overline{w}_i}$;\\
    $Append(L,\ln\lambda_{i\rm max})$;\\
 }
 $Sort(L)$;\\
 \For{each $j$ in range $2n$}{
    \For{each $i$ in range $n$}{
        \If{$L_j\leq \ln{\lambda_{i\rm min}}$}{$M_{ji}=\underline{w}_i$;}
        \If{$L_j\geq \ln{\lambda_{i\rm max}}$}{$M_{ji}=\overline{w}_i$;}
        \Else{$M_{ji}=\underline{w}_i+(\overline{w}_i-\underline{w}_i)({\ln{\frac{L_j}{a_i}}}+\ln{(\overline{w}_i-\underline{w}_i)})$;}
    }
    $m_j = Sum(M_j)$;\\
 }
 \For{each $j$ in range $2n$}{
    \If{$m_j \leq w \leq m_{j+1}$}{$break$;}
 }
 $J = j$;\\
 $Slop[J \rightarrow J+1 ]=\frac{L_{J+1}-L_J}{m_{J+1}-m_{J}}$;\\
 $\ln{\lambda_{\rm new}}=(Slop[J \rightarrow J+1 ])(w-m_J)+L_J$;\\
 \For{each $i$ in range $n$}{
    \If{$\ln{\lambda_{\rm new}}\leq \ln{\lambda_{i\rm min}}$}{$w_i=\underline{w}_i$;}\newpage
    \If{$\ln{\lambda_{\rm new}}\geq \ln{\lambda_{i\rm max}}$}{$w_i=\overline{w}_i$;}
    \Else{$w_i=\underline{w}_i+(\overline{w}_i-\underline{w}_i)(\ln{\frac{\lambda_{\rm new}}{a_i}}+\ln{(\overline{w}_i-\underline{w}_i)})$;}
    }
 $W^o=(w_1, w_2,\cdots,w_n)^T$;\\
 \For{each $i$ in range $n$}{
    $c_{i1} = c_i(W^*[i])$;\\
    $c_{i2} = c_i(W^o[i])$;\\
 }
 $C_1 = (c_{11},\cdots,c_{n1})^T, C_2 = (c_{12},\cdots,c_{n2})^T$;\\
 \For{each $t$ in range $T_2$}{
    \For{each $i$ in range $n$}{
        $C_1[i] = \sum_{j=1}^{n}A[i][j]*C_1[j]$;\\
        $C_2[i] = \sum_{j=1}^{n}A[i][j]*C_2[j]$;
    }
 }
 \If{$C_1[1]\leq C_2[1]$}{$W = W^*$;}
 \Else{$W = W^o$;}
 $return\quad W$;\\
\end{algorithm}

\subsubsection{The cost function is quadratic}

We then consider the quadratic function for $c_i(w_i)$ by following previous works~\cite{Pantoja_a12,Pantoja_a19}. To be specific, we set
\begin{eqnarray*}
c_i(w_i)&=&\int^{w_i}_0 H_i(w_i)dw_i.
\end{eqnarray*}
Here $H_i(w_i)$ is the cost function for the unit task load of individual $i$, given by
\begin{eqnarray*}
  H_{i}(w_i)&=&
  \left\{
  \begin{aligned}
  &b_i,  \quad  \text{if}\  0<w_i \leq \underline{w}_i,\\
  &a_i(w_i-\underline{w}_i)+b_i,  \quad  \text{otherwise},\\
 \end{aligned}
  \right.
\end{eqnarray*}
where $a_i$ and $b_i$ are positive coefficient factors and $b_i$ denotes the basic yield amount when individual $i$ bears the lower limit of task amount $\underline{w}_i$. Accordingly, we have
\begin{equation}\label{quadratic}
c_i(w_i)=\frac{1}{2}a_i(w_i-\underline{w}_i)^2+b_iw_i.
\end{equation}

We proceed similarly to the previously discussed case, but now the form of the cost function is defined by Eq.~\eqref{quadratic}. Thus we have
\begin{eqnarray*}
 w_i^*=\frac{\lambda+a_i\underline{w}_i-b_i}{a_i}.
\end{eqnarray*}
Furthermore, we have
\begin{eqnarray*}
\sum_{i\in \mathcal{B}}w_i^*&=&\lambda\sum_{i \in \mathcal{B}}{\frac{1}{a_i}}+\sum_{i \in \mathcal{B}}{\underline{w}_i}-\sum_{i\in\mathcal{B}}{\frac{b_i}{a_i}}.
\end{eqnarray*}
Since $w=\sum_{i=1}^nw_i^*=\sum_{i\in\mathcal{B}}w_i^*$, we have
\begin{eqnarray*}
 w=a\lambda+b,
\end{eqnarray*}
where $a=\sum_{i\in\mathcal{B}}{\frac{1}{a_i}}$ and $b=\sum_{i\in\mathcal{B}}{\underline{w}_i}-\sum_{i\in\mathcal{B}}{\frac{b_i}{a_i}}$.

This leads to a linear relationship between $\sum_{i=1}^nw_i$ and $\lambda$. Similar to the previous case, we design an algorithm for quadratic cost function, which can be used for seeking the globally optimal solution. For any undirected connected topology, the details for seeking the solution is presented in Algorithm~\ref{algo:ATA2}. Besides, we see that the computation complexity is about $o(n^2(T_1+T_2))$.

\begin{algorithm}\label{algo:ATA2}
 \caption{Algorithm of task allocation for quadratic cost function}
 \label{algo:fa}
 \SetCommentSty{small}
 \tiny
 \KwIn{Total tasks $w$, the number of individual $n$, the upper of tasks $\overline{W}=(\overline{w}_1,\cdots,\overline{w}_n)$, the lower of tasks $\underline{W}=(\underline{w}_1,\cdots,\underline{w}_n)^T$, time scale $T_1$ and $T_2$ , initial value $W^*=(w_1(0),\cdots,w_n(0))$, coefficient factor $a=(a_1,\cdots,a_n)^T$ and $b=(b_1,\cdots,b_n)^T$, $n\times n$ adjacency matrix $A=[a_{ij}]$}
 \KwOut{Allocated tasks $W=(w_1, w_2,\cdots,w_n)^T$}
 \LinesNumbered
 \While{$t<T_1$}{
    \For{each $i$ in range $n$}{
        $w_i(t+1)=w_i(t)+\Delta t(\frac{w_i(t)}{w}f_i(w_i(t))(\sum_{j\in N_i}w_j(t)
        -\sum_{j\in N_i}{f_j(w_j(t))w_j(t)}))$;
    }
    $t++$;
 }
 $W^* = (w_1(T),\cdots,w_n(T))^T$\\
 \While{$i<n+1$}{
    $\lambda_{i\rm min}=\frac{dc_i(\underline{w}_i)}{d\underline{w}_i}$;\\
    $Append(L,\lambda_{i\rm min})$;\\
    $\lambda_{i\rm max}=\frac{dc_i(\overline{w}_i)}{d\overline{w}_i}$;\\
    $Append(L,\lambda_{i\rm max})$;\\
    $i++$;
 }

 $Sort(L)$;\\
 \While{$j<2n+1$}{
    \For{each $i$ in range $n$}{
        \If{$L_j\leq {\lambda_{i\rm min}}$}{$M_{ji}=\underline{w}_i$;}
        \If{$L_j\geq {\lambda_{i\rm max}}$}{$M_{ji}=\overline{w}_i$;}
        \Else{$M_{ji}=\frac{{L_j}+a_i\underline{w}_i-b_i}{a_i}$;}
    }
    $m_j = Sum(M_j)$;\\
    $j++$;
 }
 \For{each $j$ in range $2n$}{
    \If{$m_j \leq w \leq m_{j+1}$}{$break$;}
 }
 $J = j$;\\
 $Slop[J \rightarrow J+1 ]=\frac{L_{J+1}-L_J}{m_{J+1}-m_{J}}$;\\
 $ {\lambda_{\rm new}}=(Slop[J \rightarrow J+1 ])(w-m_J)+L_J$;\\
 \newpage
 \While{$i<n+1$}{
    \If{$ {\lambda_{\rm new}}\leq {\lambda_{i\rm min}}$}{$w_i=\underline{w}_i$;}
    \If{$ {\lambda_{\rm new}}\geq {\lambda_{i\rm max}}$}{$w_i=\overline{w}_i$;}
    \Else{$w_i=\frac{{\lambda_{\rm new}}+a_i\underline{w}_i-b_i}{a_i}$;}
    $i++$;
    }
 $W^o=(w_1, w_2,\cdots,w_n)^T$;\\
 \While{$i<n+1$}{
    $c_{i1} = c_i(W^*[i])$;\\
    $c_{i2} = c_i(W^o[i])$;\\
    $i++$;
 }
 $C_1 = (c_{11},\cdots,c_{n1})^T, C_2 = (c_{12},\cdots,c_{n2})^T$;\\
 \While{$t<T_2$}{
    \For{each $i$ in range $n$}{
        $C_1[i] = \sum_{j=1}^{n}A[i][j]*C_1[j]$;\\
        $C_2[i] = \sum_{j=1}^{n}A[i][j]*C_2[j]$;
    }
    $t++$;
 }
 \If{$C_1[1]\leq C_2[1]$}{$W = W^*$;}
 \Else{$W = W^o$;}
 $return\quad W$;\\
\end{algorithm}

\section{Numerical examples}
By using the above specified methods, we here provide numerical examples about searching the globally optimal solution to the limited task allocation with exponential or quadratic cost function.

\subsection{The case of $W^*\in NE(F)\cap \textit{S}$}
\subsubsection{{The cost function is exponential}}

\begin{figure}
\begin{center}
\includegraphics[width=3in]{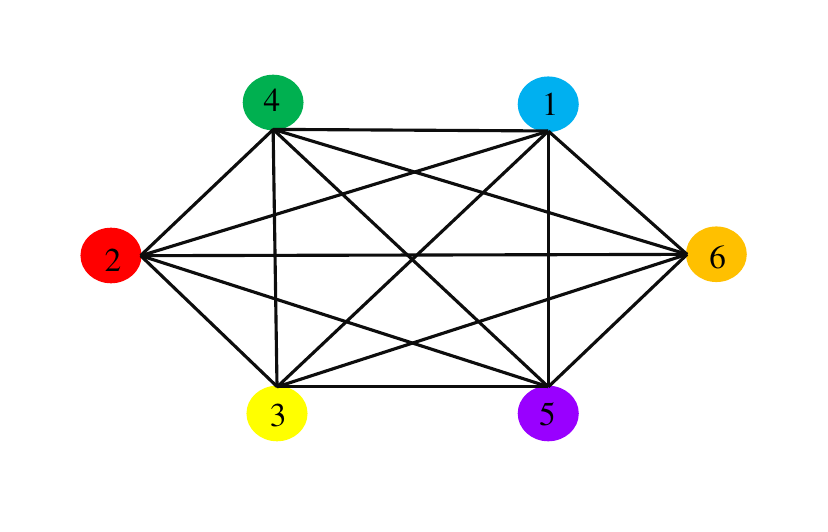}
\caption{Illustrative figure showing a connected graph of six nodes in the limited task allocation problem.
Each node represents an individual and a link between
two nodes implies that these two individuals can communicate with each other.}\label{fig:1}
\end{center}
\end{figure}

We present an example where $6$ individuals, represented by nodes on a connected graph, as shown in Fig.~\ref{fig:1}, participate in the limited task allocation and we set that $w=2800$. Here the cost function for each individual is given Eq.~\eqref{exp}. The parameter values are set as
\begin{eqnarray*}
\underline{W}&=&(\underline{w}_1,\underline{w}_2,\cdots,\underline{w}_6)^T=(0,0,0,0,0,0)^T,
\end{eqnarray*}
\begin{eqnarray*}
\overline{W}&=&(\overline{w}_1,\overline{w}_2,\cdots,\overline{w}_6)^T\\
&=&(750,800,1400,1000,900,1700)^T,
\end{eqnarray*}
and
\begin{eqnarray*}
a &=&(a_1,a_2,\cdots,a_6)^T\\
&=&(750,800,1400,1000,900,1700)^T.
\end{eqnarray*}

\begin{figure*}
\begin{center}
\includegraphics[width=6in]{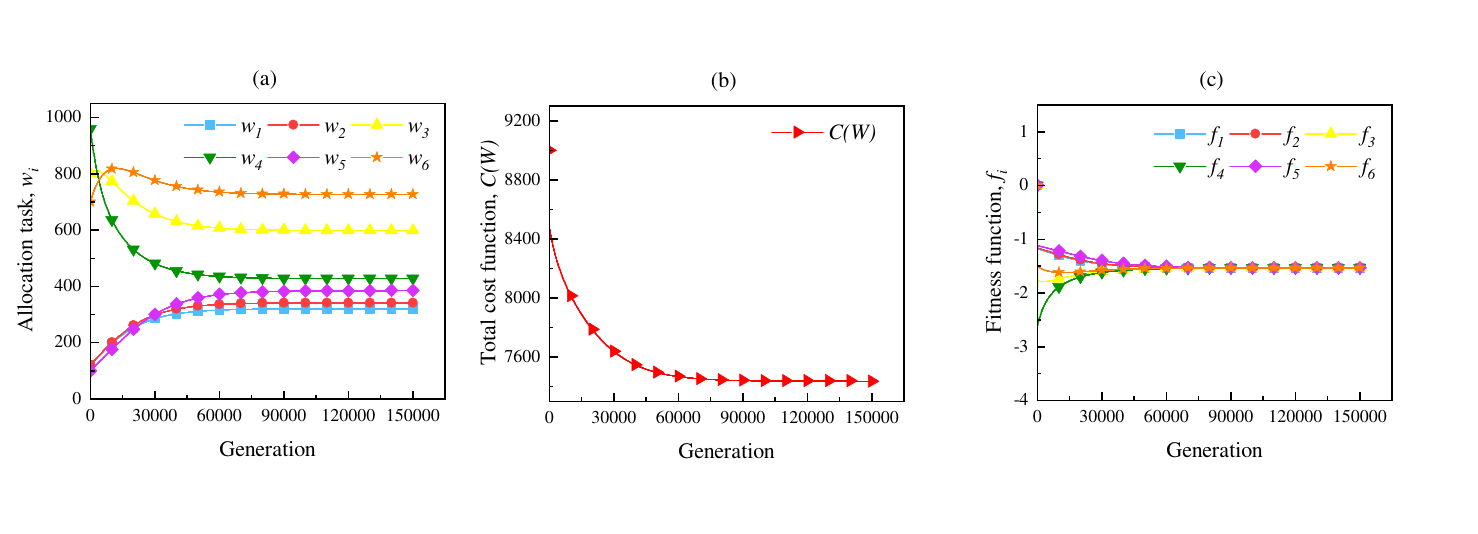}
\caption{The time evolution of allocated tasks $w_i$, the total cost function $C(W)$, and individual fitness function $f_i$. We can find that the total cost function is decreasing monotonically as time increases and finally reaches the minimal value. Correspondingly, the allocation task for each individual gradually converges to the optimal solution, and meanwhile each individual has the identical fitness value. Parameter values: $n=6, w=2800$, and $\Delta t=0.0001$. The cost function is exponential.
}\label{fig:2}
\end{center}
\end{figure*}

Furthermore, based on Eq.~\eqref{key11}, we present the evolution of allocated tasks $w_i$, the total cost function $C(W)$, and fitness function $f_i$ as a function of time, as shown in Fig.~\ref{fig:2}. We can find that the total cost function 
is decreasing monotonically as time increases and finally reaches the minimal value. Correspondingly, the allocation task for each individual gradually converges to the optimal solution, and meanwhile each individual has the identical fitness value.

\subsubsection{{The cost function is quadratic}}
By using the same multi-agent system as specified in Fig.~\ref{fig:1}, we consider that the cost function for each individual is given as defined by Eq.~\eqref{quadratic}. Here the parameter values are set as
\begin{eqnarray*}
\underline{W}&=&(\underline{w}_1,\underline{w}_2,\cdots,\underline{w}_6)^T=(700,350,200,50,40,200)^T,
\end{eqnarray*}
\begin{eqnarray*}
\overline{W}&=&(\overline{w}_1,\overline{w}_2,\cdots,\overline{w}_6)^T=(980, 580, 350, 170, 150, 790)^T,
\end{eqnarray*}
\begin{eqnarray*}
a &=&(a_1,a_2,\cdots,a_6)^T\\
&=&(0.006,0.008,0.01,0.012,0.0132,0.00136)^T,
\end{eqnarray*}
and
\begin{eqnarray*}
b &=&(b_1,b_2,\cdots,b_6)^T=(0.4,0.2,0.5,0.56,0.828,0.88)^T.
\end{eqnarray*}

\begin{figure*}
\begin{center}
\includegraphics[width=6in]{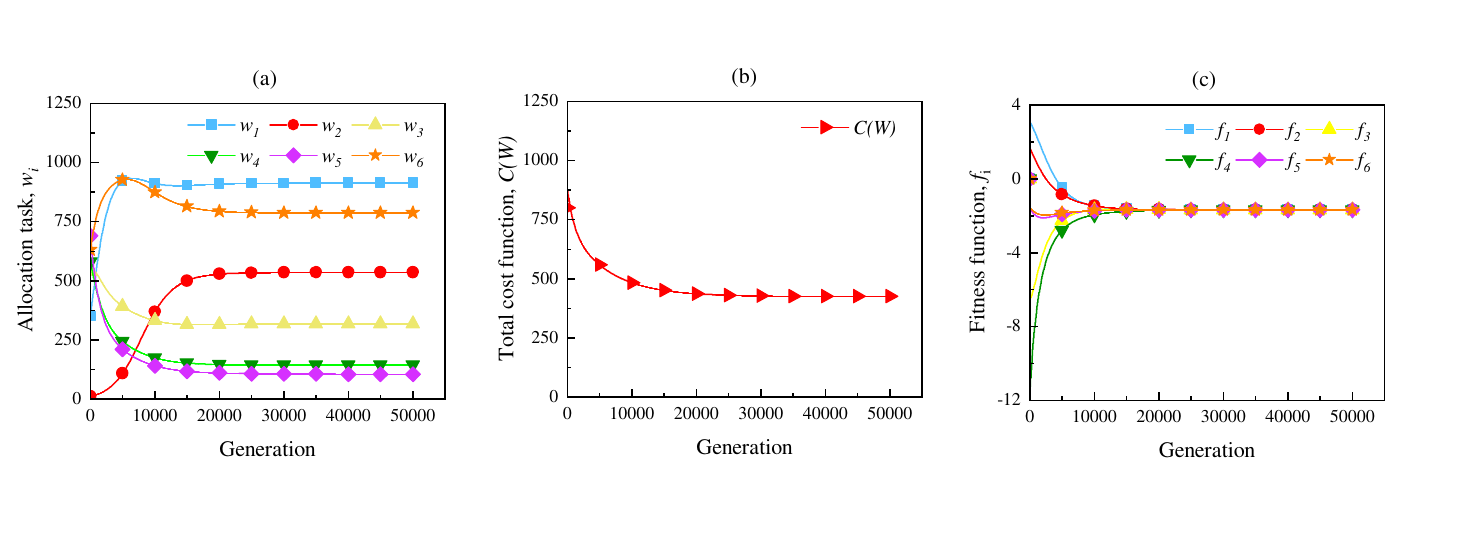}
\caption{The time evolution of allocated tasks $w_i$, the total cost function $C(W)$, and individual fitness function $f_i$. We find the total cost function monotonically decreasing as time increases and finally reaches the minimal value. Besides, the allocation task for each individual gradually converges to the optimal solution, and meanwhile each individual has the identical value of fitness. Parameter values: $n=6, w=2800$, and $\Delta t=0.001$. The cost function is quadratic.}\label{fig:3}
\end{center}
\end{figure*}

In this case the time evolution of allocated tasks $w_i$, the total cost function $C(W)$, and the individual fitness function $f_i$ are shown in Fig.~\ref{fig:3}. We can find that the total cost function monotonically decreasing as time increases and finally reaches the minimal value. Correspondingly, the allocation task for each individual gradually converges to the optimal solution, and meanwhile each individual has the identical value of fitness.

\subsection{The case of $ NE(F)\cap \textit{S}=\phi$}
\subsubsection{{The cost function is exponential}}

\begin{figure}
\begin{center}
\includegraphics[width=2in]{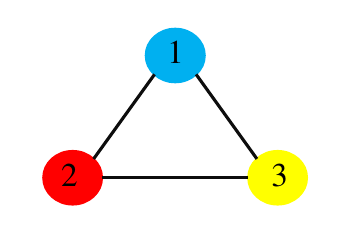}
\caption{An alternative multi-agent system for the limited task allocation problem. Nodes represent agents and links between them signal that they communicate with each other.}\label{fig:4}
\end{center}
\end{figure}

We here consider an interaction topology of three individuals as shown in Fig.~\ref{fig:4}. For the limited task allocation problem we set the total task amount $w=1150$. The cost functions for these three individuals are respectively given as
\begin{eqnarray*}
  c_1(w_1) &=& 1000e^{\frac{w_1-200}{150}} \quad(200\leq w_1\leq 350),\\
  c_2(w_2) &=& 1900e^{\frac{w_2-350}{130}}\quad(350\leq w_2\leq 480),
  \end{eqnarray*}
  and
\begin{eqnarray*}
 c_3(w_3) &=& 2300e^{\frac{w_3-410}{130}}\quad(410\leq w_3\leq 540).
\end{eqnarray*}

Based on Algorithm~1, we can respectively calculate $\ln{\lambda_{i\rm min}}$ and $\ln{\lambda_{i\rm max}}$ for each individual, and these correlation parameter values are listed in Table~I. Accordingly, we can respectively compute $m_j$ and $Slope[j \rightarrow j+1]$, where $j\in \{1,2,3,4,5,6\}$, and these correlation parameter values are listed in Table~II.
\begin{center}\label{table2}
\begin{table}
\caption{$\ln{\lambda_{i\rm min}}$ and $\ln{\lambda_{i\rm max}}$ values}
\begin{center}
  \begin{tabular}{ccccc}
\hline
$i$&$w_i$&$\ln{\lambda_i}$\\
\hline
1&$\underline{w}_{1}=200$&$\ln{\lambda_{1\rm min}}=1.897$\\
1&$\overline{w}_{1}=350$&$\ln{\lambda_{1\rm max}}=2.897$\\
2&$\underline{w}_{2}=350$&$\ln{\lambda_{2\rm min}}=2.682$\\
2&$\overline{w}_{2}=480$&$\ln{\lambda_{2\rm max}}=3.682$\\
3&$\underline{w}_{3}=410$&$\ln{\lambda_{3\rm min}}=2.873$\\
3&$\overline{w}_{3}=540$&$\ln{\lambda_{3\rm max}}=3.873$\\
\hline
\end{tabular}
\end{center}
\end{table}
\end{center}

\begin{center}\label{table2}
\begin{table}
\caption{$m_j$ and $Slope[j \rightarrow j+1]$ values}
\begin{center}
  \begin{tabular}{ccccc}
\hline
j&$L_j$
&$m_j$&$Slope[j \rightarrow j+1 ]$\\
\hline
1&1.897&960&\\
 & &     &$6.6677\times 10^{-3}$\\
2&2.682&1077.732&\\
 & &     &$3.5721\times 10^{-3}$\\
3&2.873&1131.202&\\
 & &     &$2.4388\times 10^{-3}$\\
4&2.897&1141.043&\\
 & &     &$3.8460\times 10^{-3}$\\
5&3.682&1345.153&\\
 & &     &$7.6870\times 10^{-3}$\\
6&3.873&1370&\\
\hline
\end{tabular}
\end{center}
\end{table}
\end{center}
Furthermore, the globally optimal solution in our example can be obtained, given as
\begin{center}
  $w_1=350$, $w_2=382.4$, and $w_3=417.6.$
\end{center}

\begin{figure}
\begin{center}
\includegraphics[width=4in]{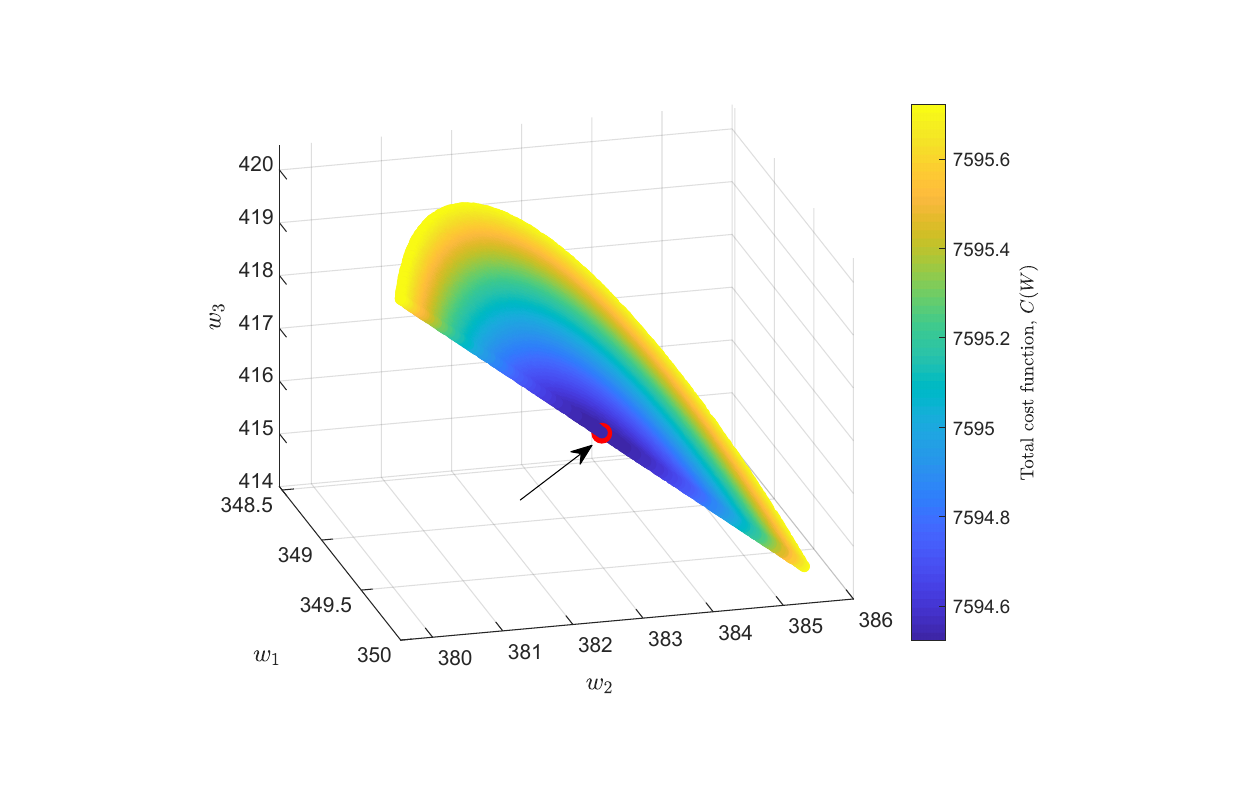}
\caption{Monte Carlo simulation results for the limited task allocation problem with exponential cost function. We randomly choose $10^8$ points in the area of feasible solutions and calculate the corresponding $C(W)$ values. These values are color coded. The red circle represents the globally optimal solution obtained numerically by using Algorithm~1. The black arrow indicates the globally optimal solution solved by Monte Carlo simulations.}\label{fig:5}
\end{center}
\end{figure}

In order to confirm the numerical outcomes obtained by Algorithm~1, we also perform Monte Carlo simulations which can be used for searching the optimal solution to the limited task allocation problem~\cite{Hartmann_ak02}. Our simulation results are shown in Fig.~\ref{fig:5} and we find that the optimal solution obtained by our Monte Carlo simulations is consistent with that obtained by Algorithm~1. The latter is indicated by a red circle in the figure.

\subsubsection{{The cost function is quadratic}}

For the same system shown in Fig.~\ref{fig:4}, we here also set that the total task amount $w=1150$ and consider that the cost functions for these three individuals are respectively given as
\begin{eqnarray*}
  c_1(w_1) &=& 0.003(w_1-200)^2+5w_1 \quad(200\leq w_1\leq 350),\\
  c_2(w_2) &=& 0.004(w_2-350)^2+5.4w_2\quad(350\leq w_2\leq 480),
  \end{eqnarray*}
  and
\begin{eqnarray*}
c_3(w_3) &=& 0.005(w_3-410)^2+5.6w_3\quad(410\leq w_3\leq 540).
\end{eqnarray*}

Based on Algorithm~2, we can then respectively calculate $\lambda_{i\rm min}$ and $\lambda_{i\rm max}$ for each individual, and these correlation parameter values are listed in Table~III. We can also calculate $m_j$ and $Slope[j \rightarrow j+1]$, where $j\in \{1,2,3,4,5,6\}$, and these correlation parameter values are listed in Table~IV.
\begin{center}
\begin{table}
\caption{$\lambda_{i\rm min}$ and $\lambda_{i\rm min}$ values}
\begin{center}
 \begin{tabular}{ccccc}
\hline
$i$&$w_i$&$\lambda_i$\\
\hline
1&$\underline{w}_{1}=200$&$\lambda_{1\rm min}=5.00$\\
1&$\overline{w}_{1}=350$&$\lambda_{1\rm max}=5.90$\\
2&$\underline{w}_{2}=350$&$\lambda_{2\rm min}=5.40$\\
2&$\overline{w}_{2}=480$&$\lambda_{2\rm max}=6.44$\\
3&$\underline{w}_{3}=410$&$\lambda_{3\rm min}=5.60$\\
3&$\overline{w}_{3}=540$&$\lambda_{3\rm max}=6.90$\\
\hline
\end{tabular}
\end{center}
\end{table}
\end{center}

\begin{center}
\begin{table}
\caption{$m_j$ and $Slope[j \rightarrow j+1]$ values}
\begin{center}
 \begin{tabular}{ccccc}
\hline
j&$L_j$
&$m_j$&$Slope[j \rightarrow j+1 ]$\\
\hline
1&5&960&\\
 & &     &$6\times 10^{-3}$\\
2&5.4&1026.667&\\
 & &     &$3.4286\times 10^{-3}$\\
3&5.6&1085&\\
 & &     &$2.5532\times 10^{-3}$\\
4&5.9&1202.5&\\
 & &     &$4.4444\times 10^{-3}$\\
5&6.44&1324&\\
 & &     &$10\times 10^{-3}$\\
6&6.9&1370&\\
\hline
\end{tabular}
\end{center}
\end{table}
\end{center}

In this example, the globally optimal solution is given as
\begin{center}
 $w_1=327.7$, $w_2=395.7$, and $w_3=426.6.$
\end{center}

\begin{figure}[!t]
\begin{center}
\includegraphics[width=4in]{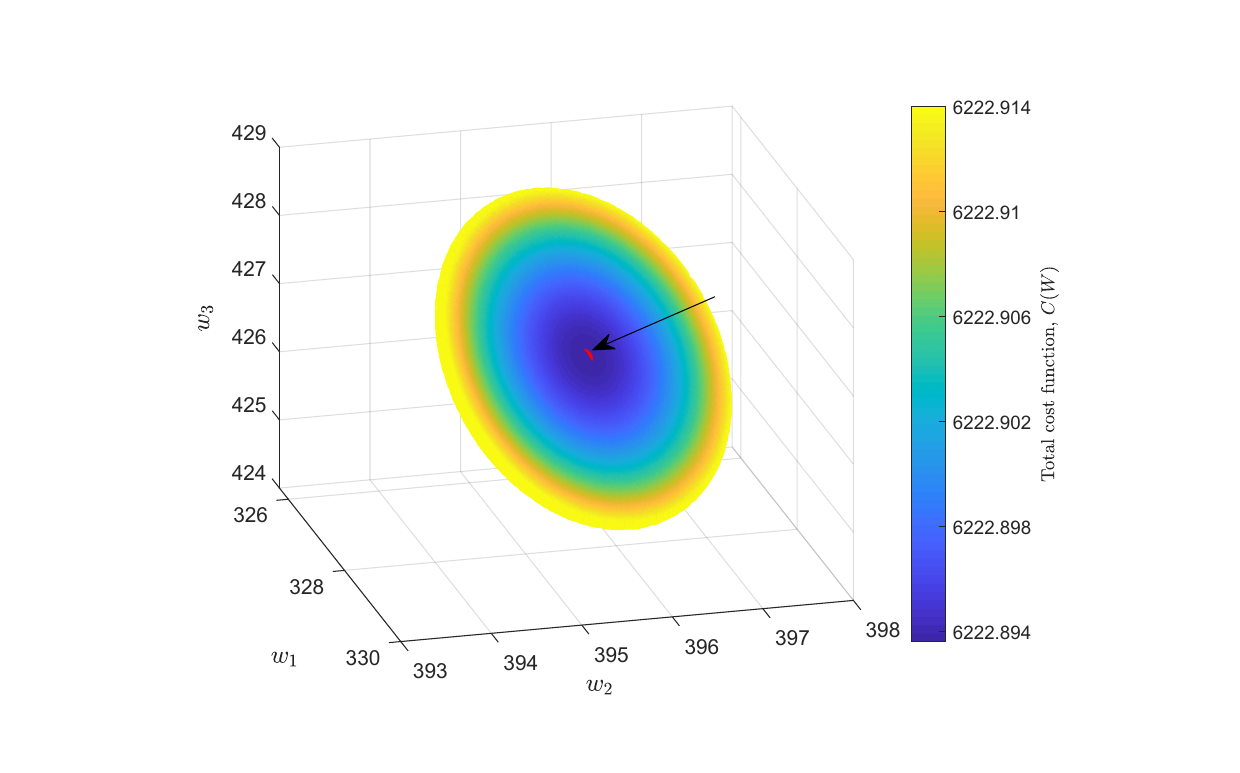}
\caption{Monte Carlo simulation results for the limited task allocation problem with the quadratic function. We randomly choose $10^8$ points in the area of feasible solutions for the limited task allocation problem and calculate the corresponding $C(W)$ values. These values are color coded, as shown in the legend. The red circle represents the globally optimal solution obtained numerically by using Algorithm~2. The black arrow indicates the globally optimal solution solved by Monte Carlo simulations.}\label{fig:6}
\end{center}
\end{figure}

The predictions of Algorithm~2 can also be checked by Monte Carlo simulations. Indeed, the latter confirm the previously obtained solution as it is shown in Fig.~\ref{fig:6}.

\section{Conclusions}

By using a game-theoretical approach, our present work has focused on the limited task allocation problem when there are individual constraints and convex cost function. We assume that each individual can afford different amounts of task. We further construct a potential game to investigate the problem in the framework of population games and accordingly calculate the fitness function for each individual in the potential game. When the Nash equilibrium point in the potential game is in the feasible solutions for the limited task allocation problem, we prove that the Nash equilibrium point is the unique globally optimal solution and is also globally asymptotically stable under DRD. Whereas when the Nash equilibrium point in the potentia game is not in the feasible solutions, we also derive the globally optimal solution and we prove that this solution is the unique solution to the corresponding limited task allocation problem. Finally, we provide some examples and perform numerical calculations by means of our proposed algorithms and DRD. We find that our numerical results also confirmed by Monte Carlo simulations support our theoretical analysis.

In this work, we respectively consider the exponential and quadratic cost functions for each individual. We stress that these two cost functions are typical convex functions, which can be found to exist in realistic situations. Furthermore, we present two algorithms to numerically seek the optimal solution to the limited task problems with the mentioned cost functions under any undirected connected topology. We believe that our work could be potential for the practical applications of collective robotics in a gaming environment~\cite{Jaleel_h20,Marden-ARCRAS18}. In addition, we study the limited task allocation problem with individual constraints under an undirected connected topology. In some situations, the interaction between individuals could be unidirectional. Therefore, it could be interesting to explore the optimal solution to the limited task allocation problem with individual constraints under a directed connected topology. This would be a research path for further work.

\section*{Acknowledgments}
This research was supported by the National Natural Science Foundation of China (Grant Nos. 61976048, 62036002, and 62103169), and the National Research, Development and Innovation Office (NKFIH), Hungary under Grant No. K142948.

\end{document}